\documentclass[pre,twocolumn,showpacs,superscriptaddress,aps]{revtex4}

\usepackage{amsfonts}
\usepackage{amsmath}
\usepackage{amssymb}
\usepackage{graphicx}
\usepackage{dcolumn}
\usepackage{dsfont}
\usepackage{times}

\begin{document}

\title{Matter-wave bright solitons in spin-orbit coupled Bose-Einstein condensates}
\author{V. Achilleos}
\affiliation{Department of Physics, University of Athens, Panepistimiopolis,
Zografos, Athens 15784, Greece}
\author{D. J. Frantzeskakis}
\affiliation{Department of Physics, University of Athens, Panepistimiopolis,
Zografos, Athens 15784, Greece}
\author{P. G. Kevrekidis}
\affiliation{Department of Mathematics and Statistics, University of Massachusetts,
Amherst MA 01003-4515, USA}
\author{D. E. Pelinovsky}
\affiliation{Department of Mathematics, McMaster University, Hamilton, Ontario, Canada, L8S 4K1}

\begin{abstract}
We study matter-wave bright solitons in spin-orbit (SO) coupled Bose-Einstein condensates (BECs)
with attractive interatomic interactions. We use a multiscale expansion method to identify solution families
for chemical potentials in the semi-infinite gap of the linear energy spectrum. Depending on the linear and
spin-orbit coupling strengths, the solitons may resemble either
standard bright nonlinear Schr\"{o}dinger solitons or exhibit a
modulated density profile, reminiscent of the stripe phase of SO-coupled repulsive BECs.
Our numerical results are in excellent agreement with our analytical findings, and demonstrate the
potential robustness of such solitons for experimentally relevant conditions through
stability analysis and direct numerical simulations.
\end{abstract}

\pacs{05.45.Yv, 03.75.Lm, 03.75.Mn}

\maketitle

{\it Introduction.} Gauge fields are ubiquitous in physics, as they are relevant to the interactions of
charged particles with electromagnetic fields \cite{saku} or to fundamental interactions
in elementary particle physics \cite{pes}. Ultracold atomic gases are considered as an excellent candidate
where a variety of artificial gauge fields can be realized;
see Ref.~\cite{dalib} for a review. Such gauge fields
have been recently studied in experiments \cite{spiel2,spiel1}
with binary Bose-Einstein condensates (BECs). Importantly,
synthetic magnetic fields can produce spin-orbit (SO) interactions
in a BEC consisting of (predominantly) two hyperfine states
of $^{87}$Rb, coupled by a Raman laser \cite{spiel1}.

SO-coupled BECs with repulsive interactions have become a topic of
intense investigations.
Different studies have revealed the existence of a ``stripe phase'' (consisting of a linear combination of plane
waves)~\cite{ho} and phase transitions between it and states with a single plane wave or with
zero momentum~\cite{ho1}.
The existence of topological structures, such as vortices
with~\cite{xuhan} or without~\cite{spielpra} rotation, Skyrmions \cite{skyrm}
and Dirac monopoles \cite{dirac}, as well as self-trapped states
(solitons) of an effective
nonlinear Dirac equation (NLDE), was also illustrated \cite{santos}.
While the above studies refer to BECs with repulsive interactions,
to the best of our knowledge, SO-coupled BECs
with {\it attractive interactions} have not been studied so far. The latter, is the
theme of the present work.

As it is known, attractive BECs
can become themselves matter-wave bright solitons \cite{hulet},
i.e., self-trapped and highly localized mesoscopic quantum systems that can find a variety of applications
\cite{corn}.
Here, we demonstrate the existence,
stability and dynamics of matter-wave
bright solitons in SO-coupled attractive BECs.
In particular, starting from the corresponding mean-field model,
we consider the nonlinear waves emerging in the
semi-infinite gap of the linear spectrum. Similarly to the repulsive
interaction case of Ref.~\cite{ho1}, we
find three distinct states having:
(a) zero momentum,
(b) finite momentum, $+k_0$ or $-k_0$, and
(c) stripe densities formed by the interference of
the modes with $\pm k_0$ momentum.
We analytically identify these branches,
in very good agreement with our numerical computations, and determine their spin polarizations.
We also analyze the stability of these solutions, illustrating that branches
(a) and (c) are generically stable, while branch (b)
is stable for sufficiently small atom numbers. Hence, these newly emerging
matter-wave solitons in SO-coupled BECs may be well within experimental reach.

{\it Model.} We consider SO-coupled BECs confined in a
quasi-1D parabolic trap, with
longitudinal and transverse frequencies $\omega_x \ll \omega_{\perp}$. In this setting, and
for equal contributions of Rashba \cite{rashba} and Dresselhaus \cite{dres}
SO couplings (as in the experiment of Ref.~\cite{spiel1}),
the mean-field energy functional of the system is $E = \int_{-\infty}^{+\infty}\mathcal{E}dx$, with:
\begin{eqnarray}
\mathcal{E}\!=\! \frac{1}{2}(\mathbf{\Psi}^{\dagger} \mathcal{H}_0 \mathbf{\Psi} +
g_{11}|\psi_\uparrow|^4+g_{22}|\psi_\downarrow|^4
+2g_{12}|\psi_\uparrow|^2|\psi_\downarrow|^2),
\label{hamfull}
\end{eqnarray}
where $\mathbf{\Psi}\equiv (\psi_\uparrow\;\; \psi_\downarrow)^T$, and the condensate wavefunctions
$\psi_\uparrow$ and $\psi_\downarrow$ are related to the two pseudo-spin components of the BEC.
The single particle Hamiltonian $\mathcal{H}_0$ in Eq.~(\ref{hamfull}) reads:
\begin{eqnarray}
\mathcal{H}_0=\frac{1}{2m}(\hat{p}_x \mathds{1}-k_L\hat{\sigma}_z)^2+V_{\rm tr}(x) \mathds{1}+\Omega\hat{\sigma}_x,
\label{ham0}
\end{eqnarray}
where $\hat{p}_x=-i\hbar\partial_x$ is the momentum operator in the longitudinal direction,
$m$ is the atomic mass, $\hat{\sigma}_{x,z}$ are the usual $2\times2$ Pauli matrices,
$\mathds{1}$ is the unit matrix, $k_L$ is the wavenumber of the Raman
laser which couples the two atomic hyperfine states, $\Omega=\sqrt{2}\Omega_R$
is the strength of the Raman coupling, while
$V_{\rm tr}(x)= m\omega_x^2 x^2/2$ is the harmonic trapping potential. Finally, the effective 1D coupling constants
in Eq.~(\ref{hamfull}), $g_{ij}=2\alpha_{ij}\hbar\omega_\perp$ ($i,j=1,2$), are defined by the $s$-wave scattering
lengths $\alpha_{ij}$; for attractive interactions, $\alpha_{ij}<0$.

Let us measure length in units of the transverse harmonic oscillator length
$a_\perp=\sqrt{\hbar/(m \omega_\perp)}$, energy in units of $\hbar \omega_\perp$,
and densities in units of $2|\alpha_{11}|$; furthermore,
employing the gauge
transformation $\psi_{\uparrow,\downarrow}(x,t) \rightarrow \psi_{\uparrow,\downarrow}(x,t)\exp(-i\mu t)$,
where $\mu$ is the chemical potential, we derive from Eq.~(\ref{hamfull})
the following
dimensionless equations of motion for $\psi_{\uparrow,\downarrow}$:
\begin{eqnarray}
i\partial_t\psi_\uparrow \!\!&=&\!\! \left(-\frac{1}{2}\partial^2_x-ik_L\partial_x + V_{\rm tr}(x)
- |\psi_\uparrow |^2- \beta |\psi_\downarrow|^2 \right)\psi_\uparrow \nonumber \\
& \phantom{t} & - \mu \psi_{\uparrow} + \Omega\psi_\downarrow,
\label{GP1}\\
i\partial_t\psi_\downarrow \!\!&=&\!\! \left( -\frac{1}{2}\partial^2_x+ik_L\partial_x +V_{\rm tr}(x)
-\beta|\psi_\uparrow |^2 - \gamma |\psi_\downarrow |^2 \right)\psi_\downarrow
\nonumber \\
& \phantom{t} & - \mu \psi_{\downarrow} + \Omega\psi_\uparrow, \label{GP2}
\end{eqnarray}
where $V_{\rm tr}(x) = (\omega_x/\omega_\perp)^2 x^2/2$, $\beta=|\alpha_{12}/\alpha_{11}|$,
$\gamma=|\alpha_{22}/\alpha_{11}|$, and we have used the transformations
$k_L\rightarrow k_L/a_\perp$ and $\Omega\rightarrow \Omega \hbar\omega_\perp $.

Limiting cases of the system (\ref{GP1})-(\ref{GP2}) with $V_{\rm tr} = 0$
have been studied in a wide range of contexts. First,
in the absence of the kinetic ($\propto \partial_x^2$) and self-interaction ($|\psi_{\uparrow} |^2 \psi_{\uparrow}$,
$|\psi_{\downarrow} |^2 \psi_{\downarrow}$) terms, the above system
becomes the massive
Thirring model \cite{mtm}, which is a Lorentz-invariant completely integrable system of classical field theory,
possessing exact soliton solutions \cite{orf}. In the absence of the kinetic terms, but in the presence of
self-interactions, the same model has been studied in nonlinear optics; in this case, Eqs.~(\ref{GP1})-(\ref{GP2})
take the form of a NLDE,
which describes solitons in optical fiber gratings
\cite{aceves}. A similar NLDE was also used in the context of SO-coupled BECs \cite{santos} and
self-trapped states, in the form of gap solitons, were proposed.
Finally, a model similar to Eqs.~(\ref{GP1})-(\ref{GP2}), which includes the kinetic terms with a dispersion
coefficient $D$, was studied in Ref.~\cite{alan}; this model, which finds applications to two coupled planar
nonlinear optical waveguides, supports so-called ``embedded solitons'' for various values of $D$
(and frequency $\omega$); these solitons, however, are generally only
semi-stable.

Here, we will use a multiscale expansion method to derive approximate soliton solutions of
Eqs.~(\ref{GP1})-(\ref{GP2}) with a frequency (chemical potential) residing in the semi-infinite gap of the linear
spectrum. The solitons will be found to be stable for a wide range of experimentally relevant parameter values.
Our analytical results will be obtained
for $\gamma=1$ and $V_{\rm tr}=0$; deviations from
this choice will be investigated numerically and
they will not qualitatively alter our results.

\begin{figure}[tbp]
\includegraphics[width=9cm]{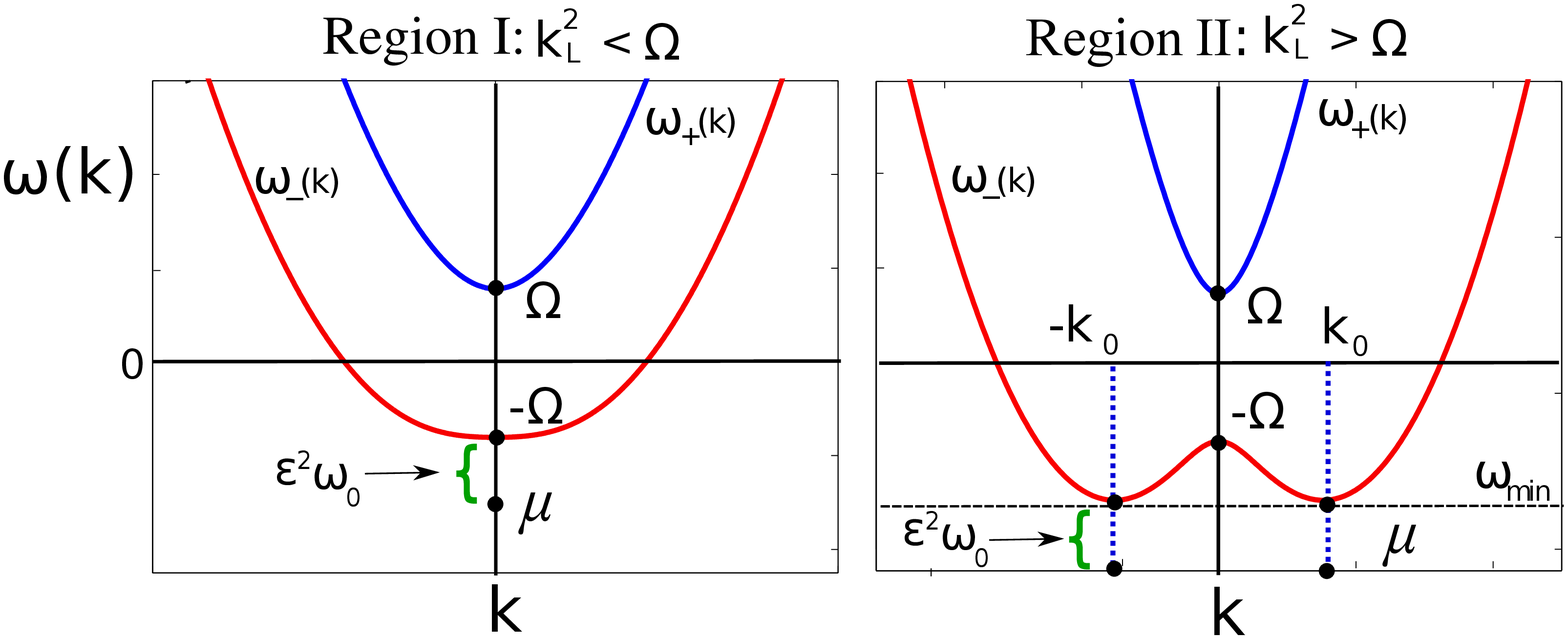}
\caption{(Color online) The linear dispersion relation (energy spectrum)
$\omega=\omega_{\pm}(k)$.
The upper branch $\omega_{+}$ has a minimum $(k, \omega)=(0, \Omega)$ in both regions I
(left panel) and II (right panel), corresponding to $k_L^2<\Omega$ and $k_L^2>\Omega$.
The lower branch $\omega_{-}$
has a minimum (maximum) $(k, \omega)=(0, -\Omega)$ in region I (region II);
in region II, there also exist two minima $(\pm k_0, \omega_{\rm min})$.
}
\label{disp1}
\end{figure}

{\it Analytical results.}
Seeking small-amplitude solutions $\propto \exp[i(kx-\omega t)]$
of Eqs.~(\ref{GP1})-(\ref{GP2}) with $\mu = 0$, we obtain the following dispersion relation (energy spectrum):
\begin{equation}
\omega_{\pm}(k) = \frac{1}{2} k^2 \pm\sqrt{k_L^2k^2+\Omega^2},
\label{dr}
\end{equation}
which features two distinct branches. The upper branch,
$\omega_{+}(k)$,
always has a minimum at $(k, \omega)=(0, +\Omega)$, and the lower branch,
$\omega_{-}(k)$,
has different behaviors depending on the sign of the parameter $\Delta \equiv 1-k_L^2/\Omega$:
if $\Delta >0$ then this branch has a minimum
$(k, \omega)=(0, -\Omega)$ (region I), while if
$\Delta <0$, $\omega_{-}(k)$ has a maximum
$(k, \omega)=(0, -\Omega)$ and two minima
$(\pm k_0, \omega_{\rm min})$ (region II).
The dispersion relation $\omega_{\pm}(k)$ is shown in Fig.~\ref{disp1}; clearly,
in the linear regime,
the lowest energy states in region I
can only have
zero momentum, $k=0$, while in region II they may have
either a positive or negative momentum, $\pm k_0$, or
they can be a linear superposition of both modes with momentum $\pm k_0$,
thus forming the ``stripe phase'' \cite{ho}.

For $\mu < -\Omega$ in region I, or $\mu < \omega_{\rm min}$ in region II, there exists a semi-infinite
gap where linear modes do not propagate. However,
matter-wave bright solitons
with energies inside the semi-infinite gap can be found analytically via a multiscale expansion method.
In particular, let $\mu = -\Omega -\epsilon^2\omega_0$ in region I and
$\mu = \omega_{\rm min} -\epsilon^2\omega_0$ in region II, where
$\epsilon$ is a formal small parameter and $\omega_0$ is a free positive parameter
(with $\omega_0/\Omega =\mathcal{O}(1)$), which sets the energy difference, $\epsilon^2 \omega_0$,
from the linear limit inside the semi-infinite gap (cf.~Fig.~\ref{disp1}).
We seek solutions of Eqs.~(\ref{GP1})-(\ref{GP2}) in the form:
\begin{eqnarray}
\left(
\begin{array}{c}
\psi_\uparrow(x,t) \\
\psi_\downarrow(x,t)
\end{array}
\right) =
\left(
\begin{array}{c}
\epsilon A(X) \\
\epsilon B(X)
\end{array}
\right) e^{iKx},
\end{eqnarray}
where $A(X)$ and $B(X)$ are unknown functions
of the slow
variable $X \equiv \epsilon x$, while the momentum $K$ is chosen as
$K=0$ in region I and $K= \pm k_0$ in region II. Expanding $A(X)$ and $B(X)$
as a series in $\epsilon$, i.e., $A(X) = \sum_{n \geq 0} \epsilon^n a_n(X)$ and
$B(X) = \sum_{n \geq 0} \epsilon^n b_n(X)$, and substituting
the above expressions in Eqs.~(\ref{GP1})-(\ref{GP2}), we obtain the following.

\begin{figure}[tbp]
\includegraphics[width=7cm]{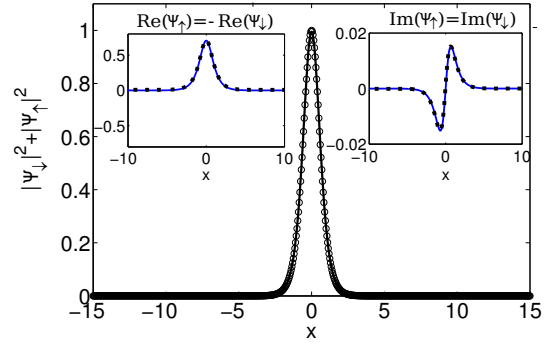}
\caption{(Color online) Density profile of the bright
soliton in region I. Solid line and circles depict, respectively, the analytical result [pertaining to Eq.~(\ref{sol1})]
and the numerically found exact solution. Left and right insets show, respectively, the real and imaginary
parts of the two wavefunctions. Parameters are: $k_L=8$, $\Omega=120$, $\beta=0.8$, and $\epsilon^2\omega_0=0.4$. }
\label{reg1}
\end{figure}

In Region I, the solvability conditions at the leading [$\mathcal{O}(1)$] and
first-order [$\mathcal{O}(\epsilon)$] approximations are satisfied if
$a_0=-b_0\equiv u(X)$ and $a_1 = b_1 = i(k_L/2\Omega)u'(X)$,
%
%
where $u(X)$ is an unknown complex function (primes denote derivatives with respect to $X$).
The latter is determined at the order $\mathcal{O}(\epsilon^2)$, where the solvability condition
is the following stationary nonlinear Schr\"{o}dinger (NLS) equation:
\begin{eqnarray}
u'' - \lambda u + \nu |u|^2 u =0,
\label{NLS1}
\end{eqnarray}
where the positive coefficients $\lambda$ and $\nu$ are given by:
$$\lambda = 2\omega_0 \Delta^{-1}, \quad  \nu = 2(1+\beta)\Delta^{-1}$$
(recall that $\Delta>0$ in region I).

In region II for $K = \pm k_0$, the solvability condition at the leading order
reads as a linear equation connecting functions $a_0(X)$ and $b_0(X)$, namely
$$
a_0 = - \Omega^{-1}k_L(k_L \mp k_0) b_0 = u(X).
$$
At the first order, we obtain a similar condition for the functions
$a_1(X)$ and $b_1(X)$, namely
$$
k_L(k_L \pm k_0) a_1(X) + \Omega b_1(X) = i(k_L \pm k_0)u'(X).
$$
Finally, at the order $\mathcal{O}(\epsilon^2)$, the solvability condition
is again a stationary NLS of the form of
Eq.~(\ref{NLS1}), but with
the coefficients $\lambda$ and $\nu$ now given by:
$$
\lambda = \frac{2\omega_0 k_L^2}{k_0^2}, \quad
\nu = \frac{2k_L(k_L \pm k_0)(k_L^4+k_L^2 k_0^2+\beta \Omega^2)}{\Omega^2 k_0^2}.
$$
\begin{figure}[tbp]
\includegraphics[width=7cm]{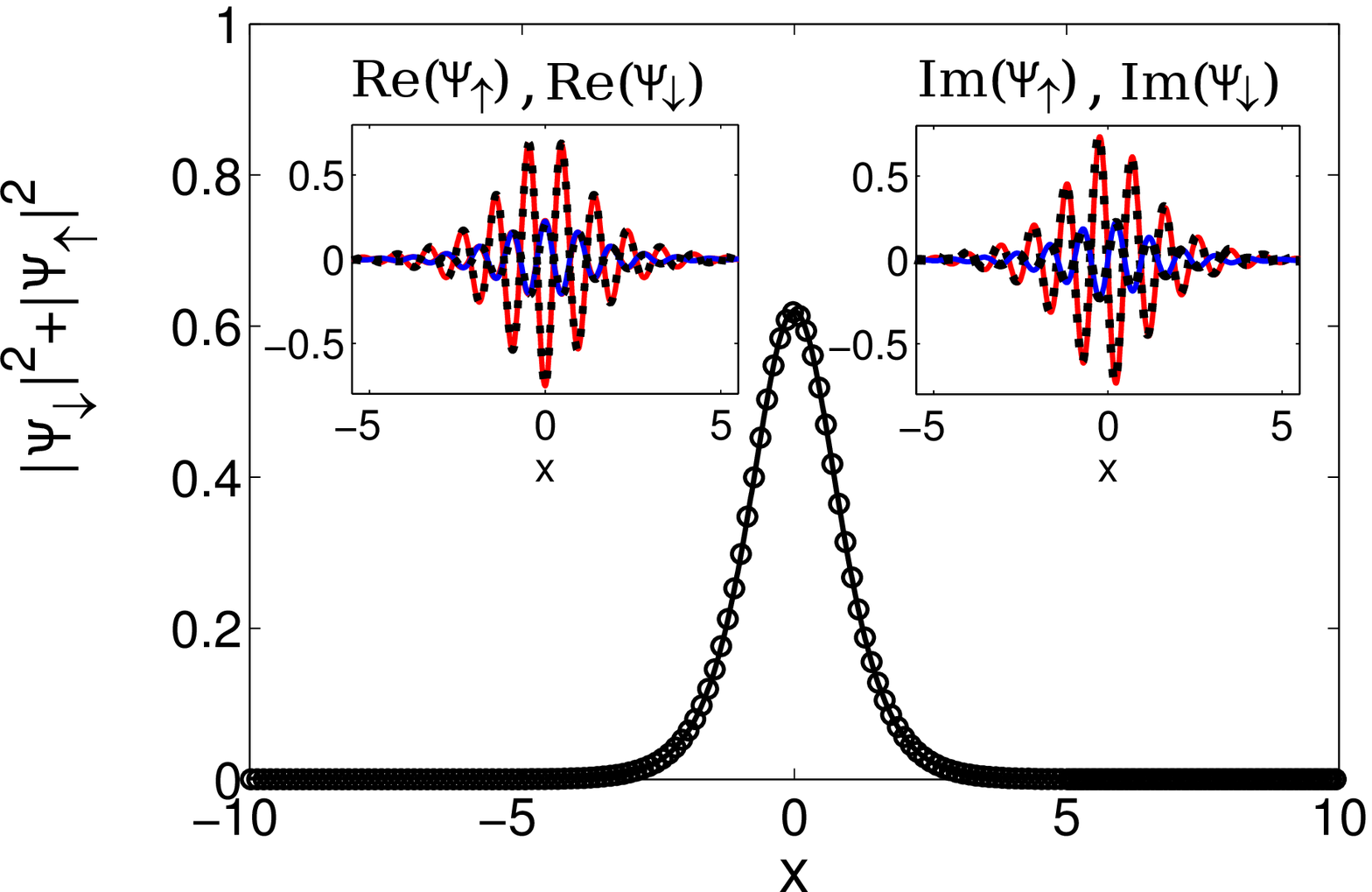}
\includegraphics[width=7cm]{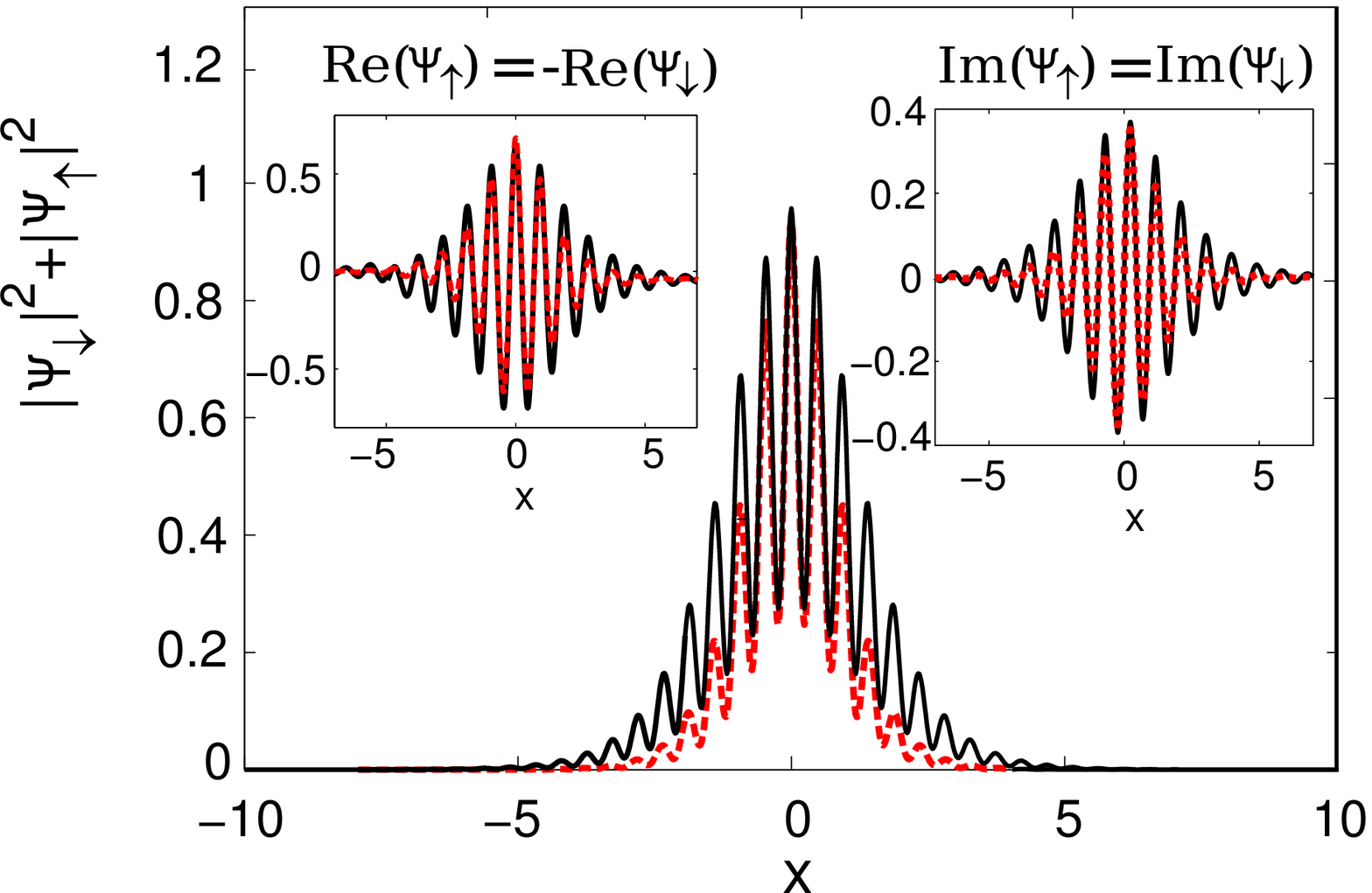}
\caption{(Color online) Same as in Fig.~\ref{reg1} but for the solitons in region II.
The top and bottom panels show, respectively, the soliton with $k=+k_0$ [cf.~Eq.~(\ref{sol2})]
and the ``stripe soliton'' [cf.~Eq.~(\ref{sol3})].
Parameters used are as in Fig.~\ref{reg1}, but with $\Omega=35$ and $\epsilon^2\omega_0=0.2$.
}
\label{reg2_ns}
\end{figure}

Taking into regard that the soliton solution of the stationary NLS Eq.~(\ref{NLS1}) is of the form
$u(X)=\sqrt{2\lambda/ \nu} \; {\rm sech} (\sqrt{\lambda}X)$, we end up with approximate
[valid up to the order $\mathcal{O}(\epsilon^2)$] soliton solutions of Eqs.~(\ref{GP1})-(\ref{GP2})
for the wavefunctions $\psi_{\uparrow,\downarrow}(x,t)$. These solutions,
characterized by the free parameter $\epsilon \sqrt{\omega_0}$ (measuring the energy difference from
the linear regime), have the following form: in region I,
\begin{eqnarray}
\!\!\!\left(
\begin{array}{c}
\psi_\uparrow \\
\psi_\downarrow
\end{array}
\right) \!&\approx&\! \epsilon \sqrt{\frac{2\omega_0}{1+\beta}}{\rm sech}\left(\epsilon \sqrt{\frac{2\omega_0}{\Delta}}x \right)
\left(\!
\begin{array}{c}
1 \\
-1
\end{array}
\!\right)\label{sol1},
\end{eqnarray}
and in region II,
\begin{eqnarray}
\!\!\!\left(
\begin{array}{c}
\psi_\uparrow \\
\psi_\downarrow
\end{array}
\right) \!&\approx&\!  \frac{\epsilon}{\sqrt{k_L \pm k_0}} f(x)
e^{\pm ik_0 x}
\left(
\begin{array}{c} \Omega \\
-k_L(k_L \pm k_0)
\end{array}
\right),\label{sol2}
\end{eqnarray}
where the function $f(x)$ is given by:
\begin{eqnarray}
\!\!\!
f(x)=\frac{\sqrt{2\omega_0 k_L}}{\sqrt{k_L^4+k_L^2 k_0^2+\beta \Omega^2}} \;
{\rm sech}\left(\epsilon \sqrt{\frac{2\omega_0 k_L^2}{k_0^2}} x\right).
\label{f}
\end{eqnarray}
Notice that
Eq.~(\ref{sol2}) describes two different soliton solutions, each corresponding to
the locations $k=\pm k_0$ of the energy minimum. We can construct still another approximate
soliton solution by using the linear combination of the above $\pm k_0$ soliton states.
In particular, Eqs.~(\ref{GP1})-(\ref{GP2}) for $\gamma = 1$ are compatible with
the symmetry $\psi_\uparrow=-\bar{\psi}_{\downarrow}$ (bar denotes complex conjugate) and
the following solution satisfies this
symmetry:
\begin{eqnarray}
\!\!\!\!\!\!\!\!\!\!\!\!
\left(
\begin{array}{c}
\psi_\uparrow \\
\psi_\downarrow
\end{array}
\right) \!\!&\approx&\!\! \epsilon f(x)
\left(
\begin{array}{c}
C_1 \cos(k_0 x) + i C_2 \sin(k_0 x) \\
-C_1 \cos(k_0 x) + i C_2 \sin(k_0 x)
\end{array}
\!
\right),
\label{sol3}
\end{eqnarray}
where $C_1 = \Omega + k_L^2$ and $C_2 = -k_0 k_L$. It is clear that, oppositely to the
solutions (\ref{sol1})-(\ref{sol2}) which have a smooth ${\rm sech}^2$-shaped density profile, the
soliton (\ref{sol3}) has a spatially modulated density profile (with a wavelength $2\pi/k_0$); thus,
this ``stripe soliton'' (\ref{sol3}) is directly analogous to the
characteristic stripe phase of
SO-coupled BECs \cite{ho,ho1},
but now for condensates with attractive interactions.
Note that only solutions (\ref{sol1}) and (\ref{sol3}) were considered in
the numerical studies of
Ref.~\cite{alan}; solution (\ref{sol2}) which does not satisfy the symmetry
$\psi_\uparrow=-\bar{\psi}_{\downarrow}$ was not previously explored.

The above solutions describe different spin polarizations of the gas: these are found
as the (normalized) longitudinal and transverse spin polarization of the solitons,
$\tilde{\sigma}_{x,z} = \langle \sigma_x \rangle/n_{\rm tot}$,
where $\langle\sigma_{x,z}\rangle \equiv \mathbf{\Psi}^{\dagger}\hat{\sigma}_{x,z}\mathbf{\Psi}$ and
$n_{\rm tot} = |\psi_\uparrow|^2+|\psi_\downarrow|^2$ is the total density.
Then, in region I, we find that
the solitons are fully polarized along the $x$-axis, i.e.,
$\tilde{\sigma}_x = -1$ (and $\tilde{\sigma}_z = 0$).
On the other hand, in region II, the stripe soliton has again $\tilde{\sigma}_z = 0$,
while the $\pm k_0$ soliton states are characterized by a finite $\tilde{\sigma}_z$, namely
$\tilde{\sigma}_z = \mp \sqrt{1-(\Omega/k_0^2)^2}$ and $\tilde{\sigma}_x = -\Omega/k_0^2$ (with
the total mean spin being $\sqrt{\tilde{\sigma}_x^2 +\tilde{\sigma}_z^2} =1$). Thus,
spin polarizations of the presented solitons bear resemblance to those
found for nonlinear
states in SO-coupled repulsive BECs \cite{ho1}.

{\it Stability and Numerical Results.} In our numerical simulations,
we have assumed a quasi-1D attractive BEC, confined in a harmonic trap with
frequencies $\omega_x=2\pi \times 20$~Hz and $\omega_\perp=2\pi \times 1000$~Hz containing approximately $10^3$ atoms,
and scattering lengths ratios $1:0.8:1$ (i.e., $\beta=0.8$); additionally, we have considered a fixed value of $k_L$,
namely $k_L=2\pi/\lambda$ with $\lambda=804$~nm and varied the parameter $\Omega$ in the range
$(1\div 10) E_L$, with $E_L= \hbar^2 k_L^2/2m$ (with $m$ being the $^7$Li mass), to identify solutions in region I or
region II (such an investigation complies with pertinent experiments with SO-coupled BECs \cite{spiel1}).
We have used a fixed-point algorithm, and an initial ansatz pertaining to solutions (\ref{sol1})-(\ref{sol2})
for regions I and II,
to find respective numerical solutions. Examples are provided in Fig.~\ref{reg1}
(for region I) and Fig.~\ref{reg2_ns} (for region II), where the density profiles,
$n_{\rm tot} = |\psi_\uparrow|^2+|\psi_\downarrow|^2$, as well
as the real and imaginary parts (insets) are shown; the analytical results (solid lines) are in excellent
agreement with the numerical ones (circles and dashed lines). Furthermore, we have numerically confirmed
(results not shown here) the existence of the presented soliton families for $\gamma \ne 1$, in a relatively wide
range of values, i.e., for $0.5 \le \gamma \le 1.5$.

\begin{figure}[tbp]
\includegraphics[width=4.2cm]{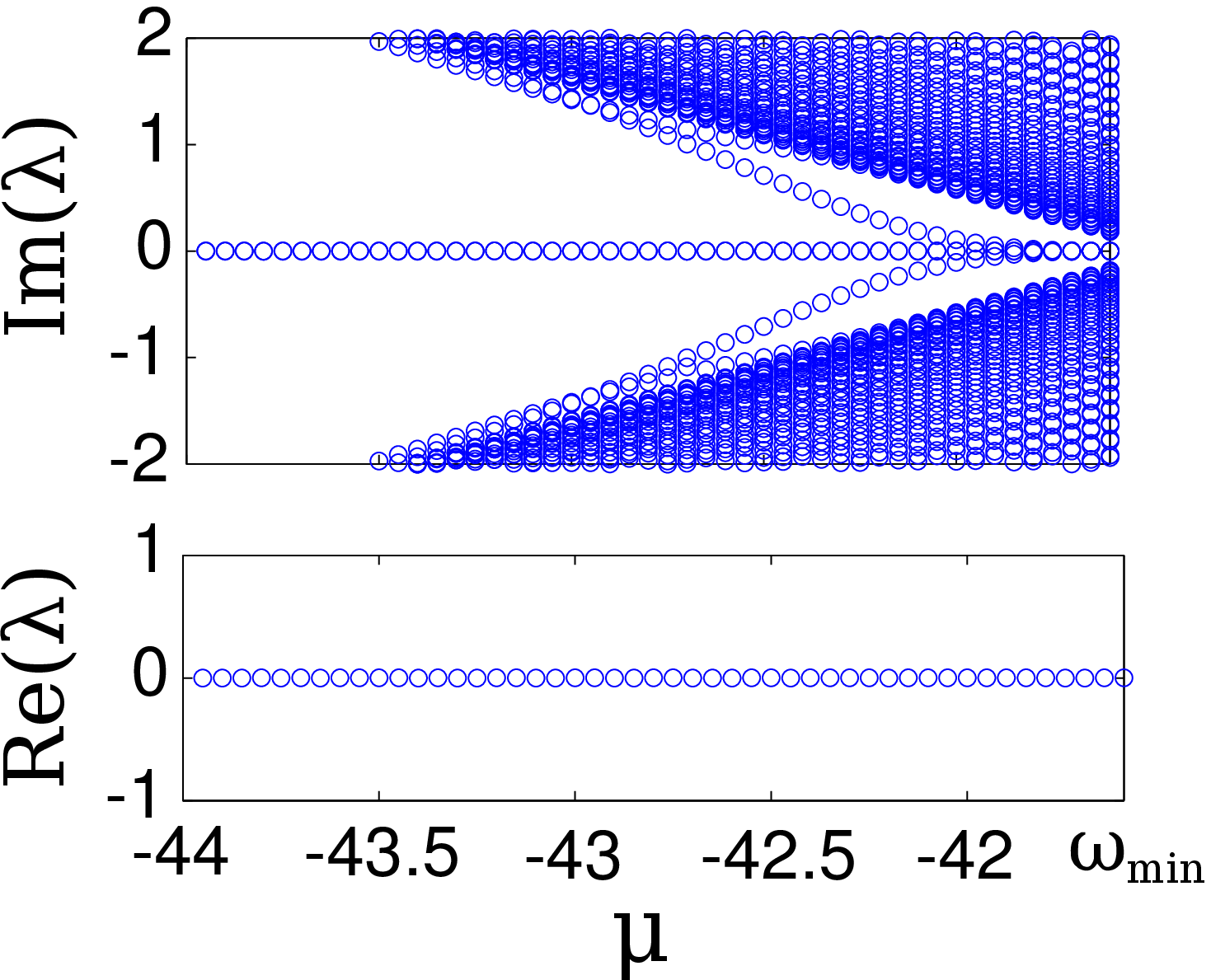}
\includegraphics[width=4.2cm]{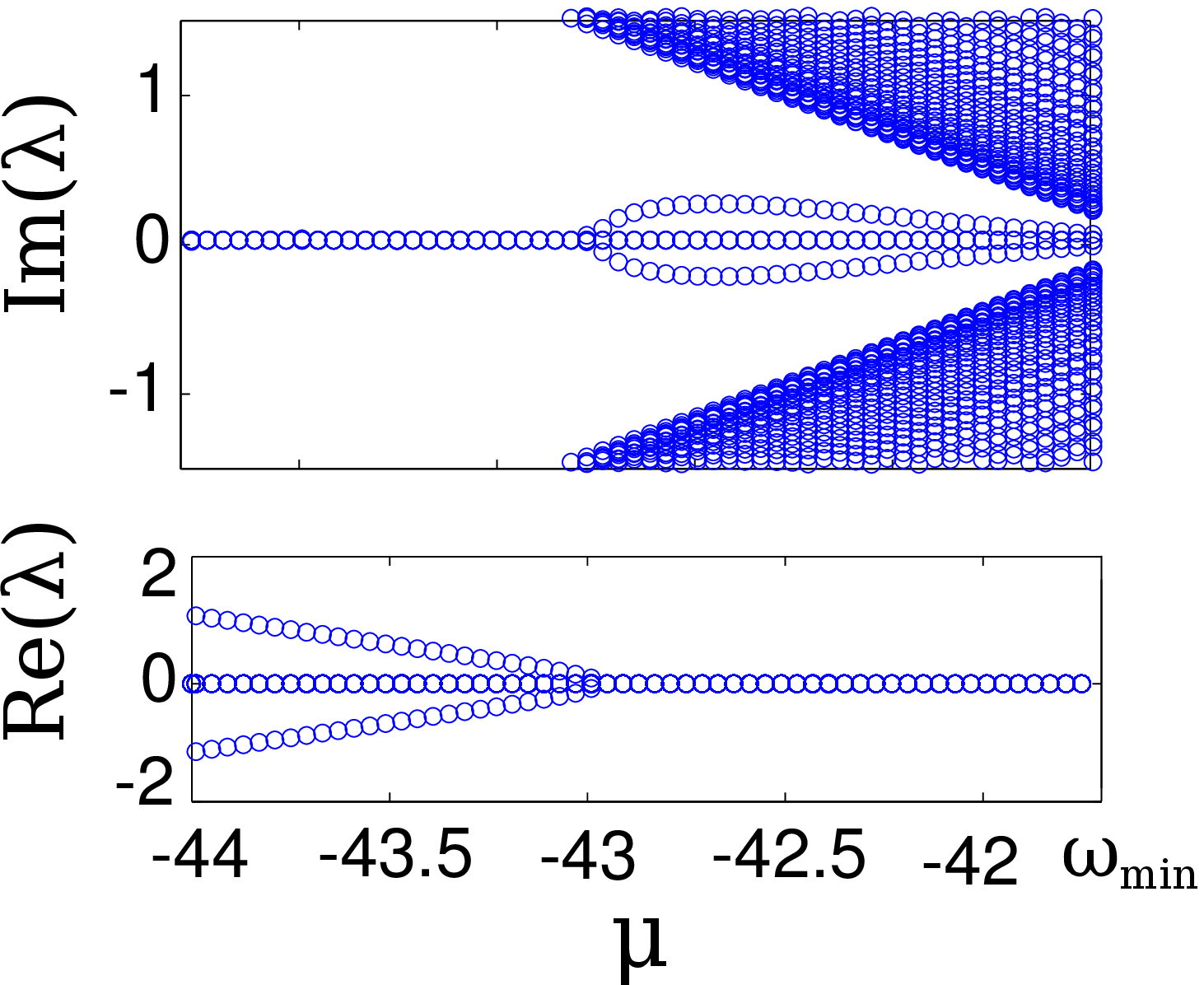}
\caption{(Color online)
Eigenvalues obtained from the spectral problem (\ref{spectrum}) for the stripe soliton (left)
and $+k_0$-soliton (right) branches. The latter becomes spectrally unstable
due to the eigenvalue pair with ${\rm Re}(\lambda) \ne 0$ for $\mu \lesssim -42.9$. Parameters are as in Fig.~\ref{reg2_ns}.
}
\label{f5}
\end{figure}

We have also studied the stability of the solitons. Because each solution family corresponds
to the energies inside the semi-infinite gap, the spectral stability of solitons
is controlled by the negative index count (explained in Ch.~4 of Ref.~\cite{Pel-book}).
Writing the spectral stability problem in the form
\begin{equation}
\label{spectrum}
H{\bf u} = i \lambda J {\bf u},
\end{equation}
where ${\bf u}$ is a $4\times 1$ vector of the perturbations
to $[\psi_{\uparrow},\bar{\psi}_{\uparrow},\psi_{\downarrow},\bar{\psi}_{\downarrow}]$,
$H$ is a $4 \times 4$ self-adjoint matrix operator associated with the right-hand-side of
Eqs.~(\ref{GP1})-(\ref{GP2}) linearized around the solitons, $J = {\rm diag}(1,-1,1,-1)$,
and $\lambda$ is a spectral parameter with the instability growth rate given by ${\rm Re}(\lambda)$
(if positive). The operator $H$ has a finite number of negative eigenvalues, denoted by $n(H)$,
and a two-dimensional kernel spanned by the symmetries of Eqs.~(\ref{GP1})-(\ref{GP2}):
\begin{equation}
\nonumber
{\bf u}_1 = [i \psi_{\uparrow},-i \bar{\psi}_{\uparrow},i \psi_{\downarrow},-i\bar{\psi}_{\downarrow}], \quad
{\bf u}_2 = \partial_x [\psi_{\uparrow},\bar{\psi}_{\uparrow},\psi_{\downarrow},\bar{\psi}_{\downarrow}].
\end{equation}
Associated with the eigenvectors of $H$, there exist generalized eigenvectors of
the spectral stability problem (\ref{spectrum}) given by solutions of the inhomogeneous equations
\begin{equation}
H {\bf v}_j = i J {\bf u}_j, \quad j = 1,2.
\label{generalized}
\end{equation}
Computing the symmetric matrix of symplectic projections with elements
$D_{ij} = \langle {\bf v}_i, iJ {\bf u}_j \rangle$ ($i,j = 1,2$),
where $\langle \cdot,\cdot \rangle$ is a standard inner product,
we denote the number of negative eigenvalues of $D$ by $n(D)$.
The negative index count is now given by $\# = n(H) - n(D)$ and this number determines
the number of unstable eigenvalues with ${\rm Re}(\lambda) > 0$
and/or the number of potentially unstable eigenvalues with ${\rm Re}(\lambda) = 0$
and negative energy in the spectral stability problem (\ref{spectrum}) \cite{Pel-book}.

To assess the stability of our solutions, we have computed indices $n(H)$ and $n(D)$
for the soliton solutions in region I and II.
For solitons in region I and the stripe solitons in region II,
we have obtained numerically that $n(H) = 1$ and $n(D) = 1$ in their existence intervals,
therefore, the negative index $\#$ is zero. This ensures
spectral stability of these solitons.
On the other hand, for $\pm k_0$-solitons in region II, we have obtained numerically that
$n(H) = 3$ in the existence interval, but $n(D)$ changes from 1 near the bifurcation at
$\mu = \omega_{\rm min}$ to 2 for smaller values of $\mu$. Therefore,
the negative index is $\# = 2$ near the
bifurcation at $\mu = \omega_{\rm min}$, due to a pair of negative energy yet
neutrally stable  eigenvalues in the spectrum of (\ref{spectrum}).
For smaller values of $\mu$, it switches to $\# = 1$ indicating a real unstable eigenvalue.

These results are confirmed by the numerical approximations of eigenvalues in
the spectral problem (\ref{spectrum}). Figure~\ref{f5} shows the
eigenvalues associated with
the stripe- and $+k_0$-solitons in region II. Spectral stability of the former is
contrasted with the potential
instability of the latter that arises when a pair of
neutrally stable eigenvalues of negative energy crosses
zero at $\mu = \mu_{\rm c} \approx -42.9$
and splits along the real axis for smaller $\mu$, yielding an exponential
growth of perturbations.

\begin{figure}[tbp]
\includegraphics[width=7cm]{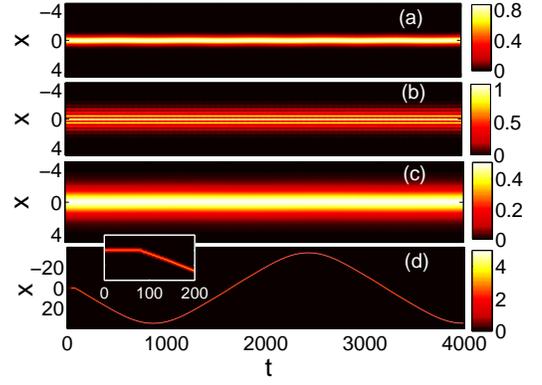}
\caption{(Color online) Contour plots showing the evolution of the total density
for solitons in region I (a) and region II (b)-(d). Panel (b)
corresponds to stripe soliton, while panels
(c) and (d) correspond to $+k_0$-solitons with $\mu=-41.77 >\mu_{\rm c}$ and
$\mu=-43 <\mu_{\rm c}$ respectively.
Other parameters are as in Fig.~\ref{reg1} (but with $\Omega=70$) and Fig.~\ref{reg2_ns}, but
now for $\gamma=0.8$ and $\omega_x/\omega_z=0.02$.
}
\label{f4}
\end{figure}

We have also
studied the soliton dynamics for
$\gamma \neq 1$ and in the presence of
the trap. We have used
our fixed point algorithm to obtain a specific soliton state;
then, the numerically found soliton was perturbed by a noise of strength $\approx 10\%$
of its initial amplitude, and the resulting state was used as initial condition for
Eqs.~(\ref{GP1})-(\ref{GP2}) with the parabolic trap.
Results of direct simulations are shown in Fig.~\ref{f4} for $\gamma=0.8$
and trap strength $\omega_x/\omega_\perp=0.02$. Solitons in region I [panel (a)], stripe solitons in
region II [panel (b)] and $k_0$-solitons with $\mu=-41.77 >\mu_{\rm c}$,
corresponding to their stability region
[panel (c)], are found to be robust
up to $t=4000$ (of the order of $1$~sec in physical units), which was the time of the simulation.
An example of unstable $k_0$-solitons with $\mu = -43 <\mu_{\rm c}$ is also illustrated [panel (d)];
in this case,
the soliton stays quiescent for small times [see inset in panel (d)], but later starts oscillating
in the trap due to the onset of the instability.

We stress that although our analytical results were obtained in the case $\gamma=1$ and $V_{\rm tr}=0$,
the simulations have revealed the existence and stability of solitons for a wide range
of values $\gamma \ne 1$, and also in the presence of the trap, as well
as for different values of $\beta$.
This clearly indicates that the presented matter-wave soliton families
have an excellent chance to be observed in experiments with SO-coupled attractive BECs.

{\it Conclusions.} In summary,
we have used a multiscale expansion method to identify matter-wave bright
soliton states
in SO-coupled BECs with attractive interactions. The solitons,
which were characterized by a chemical potential
residing in the semi-infinite gap of the linear spectrum, were found
in analytical form to
exhibit either a smooth (${\rm sech}^2$-shaped)
or a modulated density profile, strongly reminiscent of the
stripe phase of SO-coupled repulsive BECs. Our analytical predictions
were corroborated by numerical simulations, which have shown that the solitons
exist and are generally robust for a wide
range of the physical parameters involved (including chemical potential,
interatomic interaction strengths and the presence of trapping
potentials), even in the presence of noise. It would be particularly interesting
to explore higher dimensional generalizations of such solitary waves
and the potential of collapse type phenomenology~\cite{sulem} for
the various solitonic phases discussed above. Naturally, also, experimental
realizations of such SO-coupled attractive interaction BECs would
shed considerable light into such investigations.


\begin{thebibliography}{99}

\bibitem{saku} J. J. Sakurai, {\it Modern quantum mechanics} (Addison-Wesley, Redding, 1994).

\bibitem{pes} M. E. Peskin and D. V. Schroeder, {\it An introduction to quantum field theory}
(Westview Press, Boulder, 1995).

\bibitem{dalib} J. Dalibard, F. Gerbier, G. Juzeli\"{u}nas, and P. \"Ohberg, Rev. Mod. Phys.
{\bf 83}, 1523 (2011).

\bibitem{spiel2} Y.-J. Lin, R. L. Compton, K. Jimenez-Garcia, J. V. Porto, and I. B. Spielman,
Nature {\bf 462}, 628 (2009).

\bibitem{spiel1} Y.-J. Lin, K. Jimenez-Garcia, and I. B. Spielman, Nature, {\bf 471}, 83 (2011).

\bibitem{ho} T. L. Ho and S. Zhang, Phys. Rev. Lett. {\bf 107}, 150403 (2011);
S. Sinha, R. Nath, and L. Santos, Phys. Rev. Lett. {\bf 107}, 270401 (2011).

\bibitem{ho1} Y. Li, L. P. Pitaevskii, and S. Stringari, Phys. Rev. Lett. {\bf 108}, 225301 (2012).


\bibitem{xuhan} X.-Q. Xu and J. H. Han,
Phys. Rev. Lett. {\bf 107}, 200401 (2011).

\bibitem{spielpra} J. Radi\'{c}, T. A. Sedrakyan, I. B. Spielman, and V. Galitski,
Phys. Rev. A {\bf 84}, 063604 (2011);
B. Ramachandhran, B. Opanchuk, X-J. Liu, H Pu, P. D. Drummond, and H. Hu,
Phys. Rev. A {\bf 85}, 023606 (2012).

\bibitem{skyrm} T. Kawakami, T. Mizushima, M. Nitta, and K. Machida,
Phys. Rev. Lett. {\bf 109}, 015301 (2012).

\bibitem{dirac} G. J. Conduit, Phys. Rev. A {\bf 86}, 021605(R) (2012).

\bibitem{santos} M. Merkl, A. Jacob, F. E. Zimmer, P. \"{O}hberg, and L. Santos,
Phys. Rev. Lett. {\bf 104}, 073603 (2010).


\bibitem{hulet} K. E. Strecker, G. B. Partridge, A. G. Truscott and R. G. Hulet,
Nature {\bf 417}, 150 (2002);
L. Khaykovich {\it et al.},
Science {\bf 296}, 1290 (2002); S. L. Cornish, S. T. Thompson, and C. E. Wieman,
Phys. Rev. Lett. {\bf 96}, 170401 (2006).

\bibitem{corn} T. P. Billam, A. L. Marchant, S. L. Cornish, S. A. Gardiner,
and N. G. Parker, arXiv:1209.0560.

\bibitem{rashba} Y. A. Bychkov and E. I. Rashba, J. Phys. C {\bf 17}, 6039 (1984).

\bibitem{dres} G. Dresselhaus, Phys. Rev. {\bf 100}, 580 (1955).

\bibitem{mtm} W. E. Thirring, Ann. Phys. (N.Y.) {\bf 3}, 91 (1958).

\bibitem{orf} S. J. Orfanidis and R. Wang, Phys. Lett. B {\bf 57}, 281 (1975);
S. J. Chang, S. D. Ellis, and B. W. Lee, Phys. Rev. D {\bf 11}, 3572 (1975);
S. Y. Lee, T. K. Kuo, and A. Gavrielides, Phys. Rev. D {\bf 12}, 2249 (1975).

\bibitem{aceves} D. N. Christodoulides and R. I. Joseph, Phys. Rev. Lett. {\bf 62}, 1746 (1989);
A. Aceves and S. Wabnitz, Phys. Lett. A {\bf 141}, 37 (1989).

\bibitem{alan} A. R. Champneys, B. A. Malomed, and M. J. Friedman,
Phys. Rev. Lett. {\bf 80}, 4169 (1998).


\bibitem{Pel-book} D. E. Pelinovsky,
{\em Localization in periodic potentials: from Schr\"{o}dinger operators to the Gross--Pitaevskii
equation} (Cambridge University Press, Cambridge, 2011).

\bibitem{sulem} C. Sulem and P. L. Sulem,
\newblock {\it The Nonlinear Schr{\"o}dinger Equation}
(Springer-Verlag, New York, 1999).


\end{thebibliography}
\end{document}